\documentclass[12pt]{article}
\newcommand{\be}{\begin{equation}}
\newcommand{\ee}{\end{equation}}
\newcommand{\ba}{\begin{eqnarray}}
\newcommand{\ea}{\end{eqnarray}}
\begin{document}
\title{Making Sense of Singular Gauge Transformations in
$1+1$ and $2+1$ Fermion Models}
\author{%
C.D. Fosco\thanks{Investigador CONICET}\\ {\normalsize\it
  Centro At\'omico Bariloche}\\
{\normalsize\it
  8400 Bariloche, Argentina}\\
\rule{0cm}{1.cm}
F.A. Schaposnik\thanks{Investigador CICBA}\\ {\normalsize\it
  Departamento de F\'{\i}sica, Universidad Nacional de La Plata}\\
{\normalsize\it
  C.C.~67, (1900) La Plata, Argentina}%
}
\date{\today}
\maketitle
\begin{abstract}
We study the problem of decoupling fermion fields in $1+1$ and
$2+1$ dimensions, in interaction with a gauge field, by performing
local transformations of the fermions in the functional integral.
This could always be done if singular (large) gauge
transformations were allowed, since any gauge field configuration
may be represented as a {\em singular\/} pure gauge field.
However, the effect of a singular gauge transformation of the
fermions is equivalent to the one of a regular transformation with
a non-trivial action on the spinorial indices. For example, in the
two dimensional case, singular gauge transformations lead
naturally to chiral transformations, and hence to the usual
decoupling mechanism based on Fujikawa Jacobians. In $2+1$
dimensions, using the same procedure, different transformations
emerge, which also give rise to Fujikawa Jacobians. We apply this
idea to obtain the v.e.v of the fermionic current in a background
field, in terms of the Jacobian for an infinitesimal decoupling
transformation, finding the parity violating result.
\end{abstract}
\date{}
\maketitle
\newpage
\section{Introduction}
We shall be concerned with ${\cal Z}(A)$, the generating functional
for massless fermions in the presence of an external gauge field
$A_\mu$, in $D$ (Euclidean) spacetime dimensions:
\begin{equation}\label{defz}
  {\cal Z}[A] \;=\; \int {\cal D}{\bar \psi} {\cal D}\psi \,
  \exp\left[ - \int d^D x\,
  {\bar \psi}(\not \! \partial + i \not \!\! A )\psi  \right] \;,
\end{equation}
for the cases $D=2$ and $D=3$.

We start from the observation that any non-trivial, regular,
external gauge field coupled to fermions can always be written in
terms of a pure gauge field. The price to be paid is that the
gauge field transformation associated with this pure gauge field
is necessarily a {\it singular} one. It is then   possible to
perform a {\it singular} gauge transformation rendering the
fermion action free. The procedure is of course formal, in the
sense that singularities in the fermionic transformations make
complicated to check whether the new variables are strictly free
and, moreover, whether they induce a non-trivial Jacobian at the
quantum level.

These are indeed the reasons why these transformations cannot be
naively used  to decouple fermions from an external gauge field.
We shall see, however, that there is a way to make sense of them.
The clue for this, lies in the observation that a singular gauge
transformation for the fermions can be written in terms of a
regular transformation that is not the same for all the spinor
components. Namely, this regular transformation does not change
all components of the spinor with the same phase factor and can
also mix different components.

To understand the statements above, consider the  case
of 2 space-time dimensions. It is well-known in this case that
a local
 regular chiral transformation decouples massless fermions from
the gauge field~\footnote{By "local" we mean that the transformation
relates fermionic fields at the same spacetime point.}. When performed as
a path-integral change of variables,
the decoupling chiral transformation leads to a free fermionic action
but has an associated non-trivial Jacobian\cite{f}. Its calculation
\cite{rs}  allows one to have a complete control of the decoupled theory at
the quantum level. What is less known is that there is an alternative way
of decoupling two-dimensional massless fermions and this through a
singular gauge transformation. As we shall see, one can find a connection
between both transformations, the regular chiral one  and the singular
gauge one. Chiral transformations suffer from the well-known anomaly
phenomenon,
an effect of far-reaching consequences~\cite{abc}. These effects  manifest
themselves both in the operatorial and path-integral versions of the theory.

Concerning the path-integral framework, since the transformed
action is free, a non-trivial Jacobian taking into account quantum
effects is to be expected and it should coincide with the chiral
Jacobian since both transformations left us with the same free
action.

This two-dimensional example shows that either
transformation, one that changes all spinor components with the same
(singular)
phase and the other changing  each component with
regular phases differing in signs, can be used to completely decouple
massless
fermions from an external gauge field.

Once one is convinced that {\it at least in $D=2$} dimensions the
two approaches
are equivalent, however different their derivations seem to be, one
can try to make sense of singular transformations for  $D=3$ and to
determine what kind of equivalent regular transformation emerges.
This is the main purpose of the present work. We shall see that
in $D=3$ dimensions, it is  possible to obtain non perturbative
information through  the connection of infinitesimal singular gauge
transformations with regular ones, whose Jacobian can be calculated.
In particular, using the procedure described above, we compute the v.e.v.
of the fermion current in an Abelian gauge field background in $D=3$,
finding the well-known parity violating result \cite{re}-\cite{re1}.
In the present case the parity anomaly stems from the non-triviality of
a Fujikawa Jacobian as it happens in even dimensions with the axial
anomaly and the chiral Jacobian.

The plan of the paper is as follows: we first discuss in section 2
the two-dimensional case in which the connection between singular gauge
and regular chiral transformations is transparent.
Inspired by this example, we discuss in section 3 the three dimensional
case, computing the Fujikawa Jacobian for an infinitesimal regular
transformation. We then apply this result to the evaluation of the
v.e.v. of the fermion current in the presence of an external field.

\section{The Abelian case in $D=2$}\label{deq2}
This is of course the simplest case we may consider. Since, for
massless  fermions, the fermionic determinant may be exactly
calculated, everything may be checked on the safe ground of the
exact solutions. The first step in the construction of the large
gauge transformations is similar to one used in the chiral
transformation approach, namely, the decomposition of the external
gauge field $A_\mu$ into the sum of a gradient of a scalar
$\varphi$ plus the curl of a pseudoscalar $\sigma$:
\begin{equation}\label{deca2}
  A_\mu \;=\; \partial_\mu \varphi + \epsilon_{\mu\nu}
  \, \partial_\nu \sigma \;.
\end{equation}
Of course, the only pure gauge part of $A_\mu$ is,  at this point, the
gradient
term. We can however always force the curl term to also look as a pure
gauge,
 if we define another scalar field $\rho$, such that
\begin{equation}\label{defrho}
  \epsilon_{\mu\nu}\, \partial_\nu \sigma \;=\; \partial_\mu \rho \;.
\end{equation}
Taking the curl on both sides of (\ref{defrho}), we see that
\begin{equation}\label{nocomm}
  \epsilon_{\mu\nu} \, \partial_\mu \partial_\nu \rho \;=\; -
  \partial^2 \sigma = F\;,
\end{equation}
where $F(x) \equiv \epsilon_{\mu\nu}\partial_\mu A_\nu (x)$.

 Thus, the partial derivatives do not commute when acting on
$\rho$, except in the trivial case $F \,=\,0$. We shall also see
that the singularity is of the branching point type, namely,
$\rho$ is a sort of angular variable (see (\ref{exrho}) below).
Note that the singularities are in $\rho$, not on $\partial_\mu
\rho$ which is the gauge invariant part of the gauge field, and
for which we shall assume a trivial topology.

 By using (\ref{deca2})
and (\ref{defrho}), the fermionic action may be written as
\begin{equation}\label{pgaction}
  S_F \;=\; \int d^2 x \, {\bar \psi} [ \not \! \partial + i \not
  \! \partial (\varphi + \rho) ] \psi
\end{equation}
where the fermions are coupled just to a pure gauge. This suggests
the introduction of the singular (because of $\rho$) decoupling
transformations
\begin{eqnarray}\label{dectrns}
  \psi (x) &=& \exp [ -i (\varphi(x)+\rho(x))] \chi(x)\nonumber\\
  {\bar\psi} (x) &=&  {\bar\chi}(x) \exp [ i
  (\varphi(x)+\rho(x))] \;,
\end{eqnarray}
which are the transformations we want to make sense of. Of course
we will ignore the $\varphi$-dependent part, since it is regular
and produces no anomalous Jacobian (in an invariant
regularization, which we shall assume). To understand the nature
of the singularity, let us consider again equation (\ref{defrho}),
defining $\rho$. Integrating both sides along a given curve ${\cal
C}$, with origin at a fixed point $P$ of coordinates $x_P$, and
end at the point of coordinates $x$, we see that
\begin{equation}\label{intc}
  \rho(x) - \rho(x_P) \;=\; \int_{x_P}^x d\xi_\mu \;
  \epsilon_{\mu\nu} \, \partial_\nu \sigma \;.
\end{equation}
Of course, the value of $\rho (x)$ will be, in general, dependent
on both $P$ and ${\cal C}$. We shall put the point $P$ at
infinity.

In order to avoid the existence of zero modes for the Dirac operator, we
shall assume
that
\begin{equation}\label{pindep}
  \oint_{R \to \infty} d\xi_\mu \, \epsilon_{\mu\nu} \partial_\nu \sigma
  \;=\; 0
\end{equation}
holds. (The suffix ${R \to \infty}$ means that the circulation is
to be calculated along a circle of infinite radius). Indeed, because of
eq,(\ref{deca2}),  eq.(\ref{pindep}) is equivalent to
\begin{equation}\label{pindep1}
 \oint_{R \to \infty}\, d\xi_\mu \, A_\mu \;=\; 0 \;.
\end{equation}
and, by Stoke´s theorem, this is tantamount to requiring the net
magnetic flux to vanish
\begin{equation}\label{nflux}
  \int d^2 x \; \epsilon_{\mu\nu} \partial_\mu A_\nu \;=\; 0 \;.
\end{equation}
A non-zero total flux would imply the existence of zero modes, and
the vanishing of the determinant of the Dirac operator in such a background.
For a non-trivial (although topologically trivial) $A_\mu$, the
function $\rho$ will have singularities, since the circulation of
$A_\mu$ along at least one finite closed curve will be non-zero.

That the transformations (\ref{dectrns}) are not justified should
be self-evident: the new fermionic fields will have singularitites
introduced by the transformation, which then takes the original
fields out of the initial space. To make sense of these
transformations, we first consider an infinitesimal version of the
singular part of (\ref{dectrns})
\begin{eqnarray}\label{inftrns}
  \delta \psi (x) &=&  -i \eta \,\rho(x) \psi(x)\nonumber\\
  \delta {\bar\psi} (x) &=&  i \eta \,{\bar\psi}(x)\rho(x)
\end{eqnarray}
where $\eta$ is an infinitesimal parameter.

It is important to realize at  this point that in the action
(\ref{pgaction}) it is
not $\rho$ but $\not \! \partial \rho$ what appears, and this
particular combination is regular. To see this, we just need to
use the identity
\begin{equation}\label{ident}
  \not \! \partial [ \rho (x) + i \gamma_5 \sigma (x) ] \;=\; 0
  \,
\end{equation}
which is a consequence of the two dimensional Dirac algebra (we
shall see however that a similar relation exists in $D=3$). We want
to remark that  (\ref{ident}) relates
the (regular) derivatives of $\rho$ to the (also regular)
derivatives of $\sigma$. Then one realizes that, because of
(\ref{ident}), the {\em effect\/} of a chiral transformation is
equivalent to the effect of a non-chiral large gauge
transformation. (Note that we are not saying that the singular field $\rho$
is somehow transformed into a regular field by multiplying it by
$\gamma_5$).

In this way, and just by the use of (\ref{ident}), one arrives to
an equivalent representation of the action (\ref{pgaction}),
\begin{equation}\label{pcaction}
  S_F \;=\; \int d^2 x \, {\bar \psi} [ \not \! \partial + i \not
  \! \partial\varphi + \not \! \partial \gamma_5 \sigma ] \psi
\end{equation}
which is of course the starting point of the usual decoupling by a
chiral transformation.
 We may then forget the singular transformation induced by $\rho$
in favour of the regular one generated by $-i \gamma_5 \sigma$,
which is a chiral transformation. They may differ at most in a
transformation generated by $\xi$, a solution of the free equation
\begin{equation}\label{free}
  \not \! \partial \,\xi \,=\, 0 \;.
\end{equation}
Equation (\ref{free}) implies that each component of $\xi$ has to
be either  analytic or anti-analytic on the plane. Thus they are
constants. We have learnt that the only thing we could miss by
replacing the vector singular transformation by the chiral regular
one is the Jacobian due to a general global chiral transformation:
This is a $\theta$-vacua term, which is zero because of
(\ref{nflux}).

 The regular infinitesimal transformations (corresponding to the singular
ones, eq.(\ref{inftrns}), are
\begin{eqnarray}\label{chtrns}
\delta \psi (x) &=&  - \eta \sigma(x) \gamma_5 \psi(x)\nonumber\\
  \delta {\bar\psi} (x) &=&  - \eta {\bar\psi}(x) \gamma_5 \sigma (x)
  \;.
\end{eqnarray}
Once one obtains this formula for the regular transformation, one
proceeds in the usual way to derive the Jacobian, which will of
course need regularization. Note that the parameter of the
transformation is actually the same one uses when dealing with the
standard procedure involving anomalous chiral Jacobians \cite{f}.

We conclude this section by presenting the explicit form of the
fields $\sigma$ and $\rho$, since in two dimensions their
expressions are particularly simple and illuminating~\footnote{Similar
relations appear in different contexts, see for example \cite{kovner}.}.
For $\sigma$, we have
\begin{equation}\label{exsigma}
  \sigma (x) \;=\; - \frac{1}{2\pi} \, \int d^2 y \ln [\mu |x-y|]
  F (y)
\end{equation}
where $\mu$ is a constant with the dimensions of a mass. Regarding
$\rho$, it is also immediate to see that
\begin{equation}\label{exrho}
  \rho (x) \;=\; \frac{1}{2\pi} \, \int d^2 y \Theta (x-y)
  F (y)
\end{equation}
where $\Theta (x) = {\rm arg} ( \frac{x_2}{x_1} )$. It is obvious
that $\rho$ has discontinuities, associated with the branching
point of the angle function $\Theta$, while $\sigma$ does not
suffer from those singularities. Equations (\ref{exrho}) and
(\ref{exsigma}) can be collected into a single matrix equation
\begin{equation}\label{meq}
 - \sigma(x) + i \gamma_5\rho (x) \;=\; \frac{1}{2\pi} \, \int d^2 y
 \ln \left[ x_1-y_1 + i \gamma_5 (x_2-y_2) \right]
  F (y) \;.
\end{equation}

Either from these relations, or even from (\ref{defrho}), we may
see that $\rho$ and $\sigma$ satisfy the Cauchy-Riemann equations,
and are then the real and imaginary parts of an analytic function:
\begin{equation}
f = \rho + i \sigma \; \; , \; \;
 \frac{\partial}{\partial{\bar z}} f \,=\,0 \;.
\end{equation}

We may rephrase the equivalence between regular chiral and large
non chiral transformations  by saying that the effect of a  gauge
transformation generated by $\rho$ is equivalent to the one of a
chiral transformation generated by the {\em dual\/} of $\rho$,
which is $\sigma$. The term "dual" is understood here in the sense
that vortex-like configurations (remember $\rho$ is angular), are
transformed into "Coulomb" configurations (i.e., the potential of
a charge distribution).

Of course, this picture also holds the other way around: a
singular chiral pure gauge field is equivalent to a regular
non-chiral pure gauge field, and thus it can be gauged away! The
fermions are free in this case. The reason for the "asymmetry"
between these case and the original one is that one usually
regulates the Jacobian using the operator $\not \!\! D$. A more
general choice would put both cases on a similar footing.

\section{The Abelian case in $d=3$}\label{deq3}

We consider again the generating functional (\ref{defz}), but for
the $D=3$  case. Now the (Hermitian) $\gamma$ matrices
satisfy the relations
\begin{equation}
\gamma_\mu \,\gamma_\nu \,=\, \delta_{\mu\nu} I \,+ \, i \,
\epsilon_{\mu\nu\lambda} \gamma_\lambda \;,
\end{equation}
where $I$ denotes the identity matrix.

The decomposition of $A_\mu$ is now slightly different, since a
general $A$ configuration will have three independent components.
We shall use a scalar field $\varphi$ and a pseudovector
$\sigma_\mu$,
\begin{equation}\label{deca3}
  A_\mu \,=\, \partial_\mu \varphi \,+\,\epsilon_{\mu\nu\lambda}
\partial_\nu \sigma_\lambda \;.
\end{equation}
The decomposition (\ref{deca3}) apparently includes $1+3\,=\,4$
components on the right hand side. Note, however, that $A_\mu$ is
insensitive to the transformations
\begin{equation}\label{gt}
  \sigma_\lambda \; \to \; \sigma_\lambda \,+\, \partial_\lambda
  \alpha \;,
\end{equation}
which reduce by one the actual number of components. We can in
fact impose a gauge condition on $\sigma_\mu$, the natural choice
being the Lorentz condition
\begin{equation}
\partial \cdot \sigma \,=\, 0 \;,
\end{equation}
which we shall adopt. We may again write the curl term in
(\ref{deca3}) as the gradient of a scalar field $\rho$,
\begin{equation}\label{defrho3}
  \epsilon_{\mu\nu\lambda} \partial_\nu \sigma_\lambda \;=\;
  \partial_\mu \rho \;.
\end{equation}
Of course $\rho$ will have singularities
\begin{equation}\label{nocomm3}
  \epsilon_{\lambda\mu\nu} \partial_\mu \partial_\nu \rho
 \;=\; \partial_\mu \partial \cdot \sigma - \partial^2 \sigma_\mu
 \;=\; - \partial^2 \sigma_\lambda \;=\; {\tilde F}_\lambda
\end{equation}
whenever ${\tilde F}_\mu \equiv
\epsilon_{\mu\nu\lambda}\partial_\nu A_\lambda \neq 0$.
 Integrating along a curve starting at a point $P$ at
infinity, we may write
\begin{equation}\label{intc3}
  \rho (x) - \rho (x_P) \;=\; \int_{x_P}^x d\xi_\mu \,
  \epsilon_{\mu\nu\lambda} \partial_\nu \sigma_\lambda \;.
\end{equation}
The fermionic action would then be
\begin{equation}\label{pgaction3}
  S_F \;=\; \int d^3 x \, {\bar \psi} [ \not \! \partial + i \not
  \! \partial (\varphi + \rho) ] \psi
\end{equation}
where the fermions are coupled to a (singular) pure gauge field.
The singular decoupling transformations are then formally the same
as for the two-dimensional case (\ref{dectrns})
\begin{eqnarray}\label{dectrns3}
  \psi (x) &=& \exp [ -i \rho(x)] \chi(x)\nonumber\\
  {\bar\psi} (x) &=&  {\bar\chi}(x) \exp [ i \rho(x)] \;.
\end{eqnarray}
We then introduce regular infinitesimal transformations in exactly
the same way as in the two dimensional case. The
three dimensional analog of (\ref{ident})
\begin{equation}\label{ident3}
  \not\!\partial \left[\rho (x) + i \not \! \sigma (x) \right] \;=\;
  0 \;.
\end{equation}
Thus we are led to consider the infinitesimal {\em regular\/}
transformations
 \begin{eqnarray}\label{dec3}
  \delta \psi (x) &=&  \delta \!\not \! \sigma (x) \psi(x)\nonumber\\
  \delta {\bar\psi} (x) &=&   \, {\bar\psi}(x) \delta \! \not \!\sigma (x)
  \;.
\end{eqnarray}
which are the infinitesimal regular version of the decoupling singular
transformations
(\ref{dectrns3}). In the present $D=3$ case, we shall only perform
infinitesimal decoupling transformations, which will allow us to compute
v.e.v. of fermion currents $j^\mu = i\delta Z[A]/\delta A_\mu$.
We then need to consider a variation on $A_\mu$ in ${\cal Z} [A]$
\be
{\cal Z} [A + \delta A] =  \int {\cal D}{\bar \psi} {\cal D}\psi \,
  \exp\left[ - \int d^D x\,
  {\bar \psi}(\not \! \partial + i (\not \!\! A + \not \!\!
{\delta\!A} )\psi
    \right] \;.
\label{var}
\ee
Now, an infinitesimal change of fermion fields like (\ref{dec3})
can be used to eliminate $\delta A_\mu$ from the action. That change
will, very likely, produce a non trivial Jacobian ${\cal J}_{\delta \sigma}
[A , \delta A]$,
\be
{\cal Z} [A + \delta A] = {\cal J}_{\delta \sigma}[A , \delta A]
{\cal Z} [A]
\label{var2}
\ee
with
\be
{\cal J}_{\delta\sigma}[A , \delta A] = \exp \left(\int d^3x \,
\delta A_\mu \; {\cal G}_\mu[A] \right) \;. \label{j1} \ee
To obtain the connection between $\delta \sigma_\mu$ and $\delta A_\mu$,
note that from the relation (\ref{deca3}) between $\sigma_\mu$
and $A_\lambda$, we can write
\be
  \delta A_\mu^\perp \,=\, \,\epsilon_{\mu\nu\lambda}
\partial_\nu \delta \sigma_\lambda \;,
\label{deltas}
\ee
where $\perp$ denotes transverse components.
We shall connect the Jacobian (\ref{j1})  associated with
 a general gauge field variation $\delta A_\mu$ (i.e. including longitudinal
components) with that associated with
$\delta A_\mu^\perp$,
\be
{\cal J}[A + \delta A^\perp] = \exp \left(\int d^3x \, \delta
A_\mu \; {\cal A}_\mu[A] \right) \, . \label{j2} \ee
To this end, note that $\delta A_\mu^\perp$ can be thought of as a
$\delta A_\mu$ variation in a gauge where the corresponding
longitudinal component is fixed to zero. Including the 3 possible
different choices for that component gives then twice the general
variation we were looking for. Thus, a factor $3/2$ should be
included,
\be
 {\cal G}_\mu[A] = \frac{3}{2} {\cal A}_\mu[A] \, .
 \label{j3}
 \ee

Let us now compute the anomalous Jacobian associated with transformations
(\ref{dec3}).
Indeed,  under those transformations, the anomalous Jacobian reflecting the
change in
the  fermionic integration measure is
\begin{eqnarray}
   {\cal J}_{\delta \sigma} &=& {\cal J}_\psi {\cal J}_{\bar\psi}
\nonumber\\
  {\cal D} \psi &\to& {\cal D} \psi \, {\cal J}_\psi \nonumber\\
  {\cal D} {\bar\psi} &\to& {\cal D} {\bar\psi} \, {\cal
  J}_{\bar\psi} \,.
\end{eqnarray}
The Jacobians ${\cal J}_\psi$ and ${\cal J}_{\bar\psi}$ may be evaluated,
to the first order in $\delta \sigma_\mu$, as follows
 \begin{eqnarray}
 {\cal J}_\psi &=& \exp [ - {\rm Tr} \delta \!\not \! \sigma
 ]\nonumber\\
 {\cal J}_{\bar\psi} &=& \exp [ -  {\rm Tr} \delta\! \not \! \sigma ]
 \;,
 \label{traza}
 \end{eqnarray}
 where the trace is both functional and over the Dirac indices.
This trace, as it happens in the $D=2$ case, is ill-defined, and needs
regularization.
  In general, we shall have
\begin{equation}
\left[{\rm Tr}\, \delta\not\! \sigma \right]_{reg}\;=\; \int
d^3x\, \delta \sigma_\mu (x)\, {\cal A}_\mu^{reg} (x)
\end{equation}
where
\begin{equation}
 {\cal A}_\mu^{reg} (x)\;=\; {\rm tr}[
 \langle x | \gamma_\mu \, f({\not\!\!D},{\Lambda_i}) |x \rangle ]
\end{equation}
where $\Lambda_i$ ($i =1,2,\ldots$) are the Pauli-Villars regulator
masses , $f$ is a regulating function (which tends to $1$ for large values
of $\Lambda_i$ and regulates UV divergences), and ${\rm tr}$ denotes the
trace
over Dirac indices.  We shall choose a Pauli-Villars
regulator of the form
\be
f(\not\!\!D, \Lambda_i) = \frac{\Lambda_1 \Lambda_2}{\Lambda_2 - \Lambda_1}
\left(
\frac{1}{\not\!\!D + \Lambda_1} - \frac{1}{\not\!\! D + \Lambda_2}
\right)
\label{chocie}
\ee
where   two   regulators $\Lambda_i$ ($i=1,2$) with same sign are enough
 to ensure that infinities are eliminated
in the limit $\Lambda_i \to \infty$.
Only one term in
the trace survives in this limit,
\begin{equation}
 {\cal A}_\mu^{reg} (x) \;=\;
\pm  \frac{i}{24\pi}
\partial^2 \epsilon_{\mu\nu\lambda} \partial_\nu \sigma_\lambda
+ {\cal O}(\frac{1}{\Lambda})
\end{equation}
where the $\pm$ results from the (common) sign choice for the $\Lambda_i$'s.
This yields, for the Jacobian ${\cal J}_\psi$,
\begin{equation}
\label{regtr3}
{\cal J}_\psi \;=\; \exp \left( \pm \frac{i}{24\pi} \int d^3x
\delta \sigma_\mu \partial^2 A_\mu   \right)\;.
\end{equation}
The result for  ${\cal J}_{\bar\psi}$ is identical
so that the total  Jacobian   reads
\be
{\cal J}_{\delta \sigma} = \exp \left( \pm \frac{i}{12\pi} \int d^3x
\delta \sigma_\mu
\partial^2 A_\mu  \right) \;.
\label{toti}
\ee
We then see that there is a non-trivial Jacobian
associated with transformations (\ref{dec3}), the regular and infinitesimal
version of the singular gauge transformations (\ref{dectrns3}). In this
sense
transformations (\ref{dectrns3}) (or their regular counterparts (\ref{dec3})
are anomalous. Using eqs.(\ref{j1})-(\ref{j3}) we get from (\ref{toti})
the Jacobian associated with a $\delta A_\mu$ variation
\be
{\cal J}_{\delta A} =\exp \left( \pm \frac{i}{8\pi}  \int d^3x
\epsilon_{\mu \nu \lambda} \delta A_\mu
\partial_\nu A_\lambda  \right)
\label{totum}
\ee
From this we compute
the v.e.v. of the fermionic current in the presence of
an external gauge field, obtaining
\be
j_\mu = \mp \frac{1}{8\pi} \epsilon_{\mu \nu \lambda}
\partial_\nu A_\lambda
\label{cu}
\ee
This result coincides with the one originally obtained by Redlich
\cite{re}-\cite{re1} for the parity violating part of the fermion
current in a constant field strength background, using the
Schwinger method for evaluating the Euler-Heisenberg effective
action. In our calculation, valid for arbitrary field strength,
$j_\mu$ arises from a Fujikawa Jacobian. It should be noted that
our regularization prescription (eqs.(\ref{chocie})) neglects the
parity conserving part (which is absent in \cite{re1} because a
constant field strength is considered).

It is worth remarking that the decoupling procedure followed here
differs from the one applied in \cite{leguil}, since in that
reference the decoupling transformations are non-local. Moreover,
those non-local transformations induce non-anomalous (but
nevertheless non-trivial) Jacobians.

Also note that, both in $1+1$ and $2+1$ dimensions, a mass term
for the fermions is not invariant under a regular transformation,
what makes it difficult to relate the effects of singular and
regular transformations.

We conclude by stressing that both in $1+1$ and $2+1$ dimensions
the method presented in this work follows exactly the same steps:
one first writes a singular gauge transformation - a finite
decoupling one in $1+1$ dimensions, an infinitesimal one in $2+1$
dimensions. Then, one finds a regular equivalent  transformation
on fermions which can   be properly handled. Associated with these
transformations there are, at the quantum level, anomalous
Jacobians which, for the regular transformations,  can be easily
computed. The Jacobians, associated with the chiral anomaly in
$1+1$ dimensions and the parity anomaly in $2+1$ dimensions,
correctly describe  quantum aspects in both cases. This provides
an alternative way of  understanding exactly soluble two
dimensional fermion models but also of computing relevant v.e.v.'s
in higher dimensions. In fact, the same procedure described here
could be applied to the calculation of the v.e.v. of the axial
current in $3+1$ dimensions (through a decoupling transformation
analysis, which shall involve the chiral anomaly in $3+1$
dimensions), the v.e.v. of the fermion current in $4+1$ dimensions
(related to the $4+1$ dimensional parity anomaly), etc, with the
advantage that the procedure is systematic and relies on the
calculation of anomalous Jacobians whose regularization is
well-understood.

~

~
\subsection*{Acknowledgements} C.D.F. is  supported by CONICET, Argentina.
F.A.S. is supported by CICBA, Argentina
This work is  supported in part by grants from CICBA, CONICET (PIP 4330/96),
ANPCYT
(PICT 97/2285, 97/0053) and Fundaci\'on Antorchas, Argentina.


\end{document}